\documentclass[english]{article}
\usepackage[T1]{fontenc}
\usepackage[latin9]{inputenc}
\usepackage{babel}
\usepackage{url}
\usepackage{bm}
\usepackage{amsmath}
\usepackage{amssymb}
\usepackage{stmaryrd}
\usepackage{graphicx}
\usepackage[unicode=true,pdfusetitle,
 bookmarks=true,bookmarksnumbered=false,bookmarksopen=false,
 breaklinks=false,pdfborder={0 0 1},backref=false,colorlinks=false]
 {hyperref}

\makeatletter

\providecommand{\tabularnewline}{\\}

\usepackage{PRIMEarxiv}

\@ifundefined{showcaptionsetup}{}{%
 \PassOptionsToPackage{caption=false}{subfig}}
\usepackage{subfig}
\makeatother

\usepackage[style=numeric]{biblatex}
\begin{document}
\title{Locality-constrained autoregressive cum conditional normalizing flow
for lattice field theory simulations}
\author{Dinesh P. R.\thanks{Project Associate, Indian Institute of Science Education and Research,
Pune. Email: dinesh.pr@students.iiserpune.ac.in}}
\maketitle
\begin{abstract}
Normalizing flow-based sampling methods have been successful in tackling
computational challenges traditionally associated with simulating
lattice quantum field theories. Further works have incorporated gauge
and translational invariance of the action integral in the underlying
neural networks, which have led to efficient training and inference
in those models. In this paper, we incorporate locality of the action
integral which leads to simplifications to the input domain of conditional
normalizing flows that sample constant time sub-lattices in an autoregressive
process, dubbed local-Autoregressive Conditional Normalizing Flow
(l-ACNF). We find that the autocorrelation times of l-ACNF models
outperform an equivalent normalizing flow model on the full lattice
by orders of magnitude when sampling $\phi^{4}$ theory on a 2 dimensional
lattice.
\end{abstract}

\section{Introduction}

Solving path integrals in quantum field theories for theories with
large couplings involves discretization of the underlying spacetime
as lattice and numerically sampling the fields using Markov Chain
Monte Carlo (MCMC) algorithms- referred to as lattice quantum field
theory\cite{lqft-mote-carlo}. For large lattice sizes and choices
of action parameters that lead to small lattice spacing and large
correlation lengths, MCMC methods tend to suffer from long correlation
times leading to exponentially diverging computational costs- a phenomenon
known as critical slowing down (CSD)\cite{critical-sd-wolff}. While
a few non-local update algorithms have been developed for specific
models to address CSD \cite{non-local-mc-1,non-local-mc-2}, they
cannot be applied for many key theories including quantum chromodynamics
(QCD).

In recent times, machine learning-based methods \cite{wu2019solving,wu2021unbiased}
have been explored for building generative models of statistical and
field theories on a lattice. In particular, a technique based on training
normalizing flows followed by independent Metropolis-Hastings sampling
\cite{albergo2019flow} was shown to nearly eliminate CSD while modelling
the $\phi^{4}$ theory in 2 dimensions. Subsequent works \cite{gauge-equivariant1,gauge-equivariant}
proposed ways of incorporate gauge invariance of the action integral
when modelling gauge field theories using normalizing flows. This
eliminated unnecessary degrees of freedom in the variational distribution
modeled by the flow when local $SU(N)$ symmetries are present and
reducing computational costs of training and inference of the model.
This addresses a well-known problem in machine learning traditionally
known as the curse of dimensionality\cite{curse-dimensionality} where
training time and model complexity are expected to increase exponentially
with the size (dimensionality) of input/output data. Even though deep
neural networks are shown to break this curse in many cases\cite{deepnet-cod},
this is only valid under approximation that the target function depends
on low-dimensional projections of the input\cite{geometric-dl}. This
is a pretty bad approximation for generative models of lattice field
theories when correlation lengths are proportional to the lattice
size, hence eliminating unnecessary degrees of freedom is of vital
importance. Another study \cite{multimodal-flow} explores several
ways of tackling poor convergence in normalizing flows when the underlying
distribution is multimodal- like in broken symmetry phases. It also
introduces masked convolutional networks as conditional maps of coupling
layers that enforce (broken) translational invariance and a notion
of locality with stronger local dependence within the distribution.

In this work, we propose an autoregressive process of conditional
normalizing flows which enforces locality of the action integral more
strictly. By doing so, we reduce the effective input size of the conditional
normalizing flow to $O(L^{d-1})$ from the upper bound of $L^{d}$,
hence further eliminating unnecessary degrees of freedom in the spirit
of tackling the curse of dimensionality. We dub this \emph{local-Autoregressive
Conditional Normalizing Flow} (l-ACNF). After briefly introducing
the system of interest and notations in sec. \ref{subsec:Scalar-LFT},
we analyse the dependencies of autoregressive conditional distrbutions
(introduced in sec. \ref{subsec:Autoreg}) in sec. \ref{subsec:dep-set-analysis}
in fair detail. Sec. \ref{subsec:L-ACNF-metropolis} discusses the
mathematical structure of the l-ACNF model along with the accompanying
Metropolis-Hamilton process. In sec. \ref{sec:Numerical-experiments},
we describe the numerical implementation for sampling the $\phi^{4}$
theory in a 2D lattice using l-ACNF.

\section{Mathematical background}

\subsection{Scalar lattice field theory\label{subsec:Scalar-LFT}}

The system of interest consists of a hypercubic lattice of length
$L$ in $d$ dimensions where every position is labeled using a $d$-dimensional
vector $\bm{r}\in[1,L]^{d}$. For open boundary conditions, the lattice
positions $\bm{r}\notin[1,L]^{d}$ do not exist. In case of periodic
boundary conditions, we map each component of $\bm{r}\in\mathbb{Z}^{d}$
to $r_{i}\rightarrow(r_{i}-1)\mod L+1$, . State/configuration of
the system is described using scalar field values $\phi(\bm{r})$.
The configurations obey the Boltzmann distribution:
\begin{equation}
p\left(\phi(\bm{r})\right)=e^{-S[\phi]}/Z\label{eq:boltzmann}
\end{equation}
where the \emph{action} $S[\phi]$ is a functional of the field $\phi$
and $Z$ is a normalizing constant known as partition function. For
the scalar field theory with $\phi^{4}$ interactions, the action
is given by:

\begin{equation}
S[\phi]={\displaystyle \sum_{\bm{r}\in[1,L]^{d}}}\left[\phi(\bm{r}){\displaystyle \sum_{\bar{\bm{r}}}}\boxempty(\bm{r},\bar{\bm{r}})\phi(\bar{\bm{r}})+m^{2}\phi(\bm{r})^{2}+\lambda\phi(\bm{r})^{4}\right]\label{eq:latticeFT-action}
\end{equation}
where $m$, $\lambda$ are the bare mass and coupling. The d'Alembertian
in the lattice approximation is given by:
\[
\phi(\bm{r}){\displaystyle \sum_{\bar{\bm{r}}}}\boxempty(\bm{r},\bar{\bm{r}})\phi(\bar{\bm{r}})=\sum_{\mu=1}^{d}2\phi(\bm{r})^{2}-\phi(\bm{r})\phi(\bm{r}-\hat{\mu})-\phi(\bm{r})\phi(\bm{r}+\hat{\mu})
\]
where $\phi(\bm{r})$ can take any real value.

The action $S[\phi]$ in \ref{eq:latticeFT-action} has many interesting
mathematical properties. It contains only nearest neighbour product/interaction
terms $\phi(\bm{r})\phi(\bm{r}-\hat{\mu})$ and $\phi(\bm{r})\phi(\bm{r}+\hat{\mu})$,
besides powers of $\phi(\bm{r})$. This makes the action \emph{local},
a common feature in many fundamental physical theories. The action
is also symmetric/invariant to the transformation $\phi(\bm{r})\rightarrow-\phi(\bm{r})$,
along with various discrete rotations and reflections of the lattice
(hypercubic symmetry group).

\subsection{Autoregressive process\label{subsec:Autoreg}}

The Boltzmann distribution in \ref{eq:boltzmann} depends on a large
number of random variables $L^{d}=N$ and it's usually difficult to
sample from it or evaluate the PDF directly. One way to circumvent
is to model it as a product of conditional distribtutions over individual
random variables. Let's first replace the labels of lattice positions
from vectors $\bm{r}\in[1,L]^{d}$ to an integer \emph{ordering} $k\in[1,N]$.
We'll use a particular ordering that maps from vector labels as:
\begin{equation}
k(\bm{r})=\left({\displaystyle \sum_{i=1}^{d}}(r_{i}-1)L^{i-1}\right)+1\label{eq:ordering}
\end{equation}

Using a chain rule obtained from repeated application of Bayes theorem,
we can decompose the probability into a product of $N$ conditional
probabilities, in the above ordering:
\begin{align}
p(\{\phi_{k}|k\in[1,N]\}) & =p(\phi_{1},\phi_{2}\dots\phi_{N})=p(\phi_{1})p(\phi_{2}|\phi_{1})\dots p(\phi_{N}|\phi_{N-1}\dots,\phi_{1})\nonumber \\
\implies\log p(\{\phi_{k}|k\in[1,N]\}) & =\log p(\phi_{1})+\log p(\phi_{2}|\phi_{1})+\dots\log p(\phi_{N}|\phi_{N-1}\dots,\phi_{1})\\
 & =\sum_{k\in[1,N]}\log p(\phi_{k}|\phi_{<k})\label{eq:autoregressive}
\end{align}
This is referred to as an \emph{autoregressive process}. If we can
evaluate and sample the conditional probabilities sequentially, this
would yield a sample and log-probability of the entire Boltzmann distribution.
Exact forms of $p(\phi_{k}|\phi_{<k})$ are still analytically intractable
starting from \ref{eq:boltzmann}. So approaches using this principle
usually optimize the KL-divergence between a machine learning-based
variational distribution and the unnormalized Boltzmann distribution\cite{wu2019solving}.

\subsection{$d-1$ dimensional dependency sets for local interactions\label{subsec:dep-set-analysis}}

\paragraph{Ising model and open boundary conditions}

\begin{figure}
\subfloat[The conditional probability $p(\phi_{k}|\phi_{<k})$ of the field
at red position depends only on the nearest neighbours of the positions
in $\phi_{\protect\geq k}$ (coloured red/white), within $\phi_{<k}$.
The dependency set consists only of blue positions while the black
positions within $\phi_{<k}$ can be ignored.\label{fig:open-single-pos}]{\includegraphics{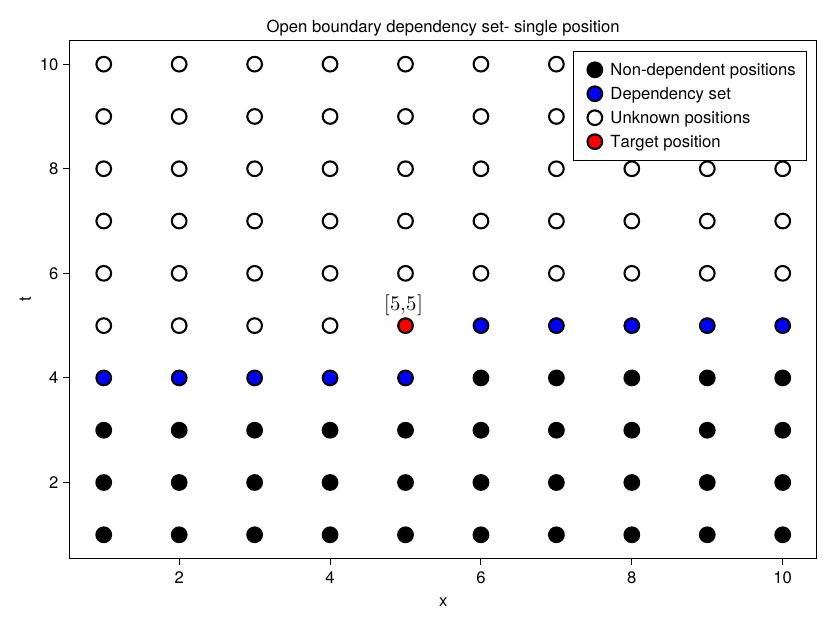}

}

\subfloat[The conditional probability of the field at red positions at a constant
time $p(\{\phi(x,t)|t=5\}|\{\phi(x,t)|t<5\})$ depends only on the
positions with $t=4$ marked blue $\{\phi(x,t)|t=4\}$ and not the
remaining ones marked black. \label{fig:open-line-pos}]{\includegraphics{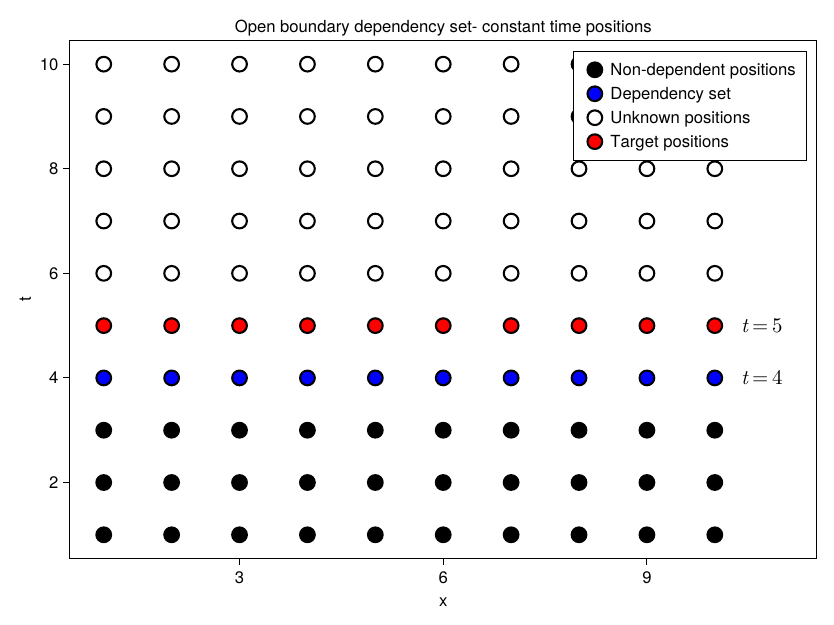}

}

\caption{Dependency sets for the conditional probabilities of field value at
single position (\ref{fig:open-single-pos}) and joint conditional
probability of positions in a constant time manifold (\ref{fig:open-line-pos}),
assuming open boundary conditions.}
\end{figure}

Examining the $k$th conditional probability $p(\phi_{k}|\phi_{<k})$
in \ref{eq:autoregressive}, its distribution in general depends on
$k-1$ values in $\phi_{<k}=\{\phi_{k-1},\dots\phi_{1}\}$. This means
the complexity of these distributions can explode if the number of
lattice points $N$ is large, which is typically the case of interest.
However for systems with nearest neighbour interactions, the \emph{dependency
set} is significantly smaller \cite{pr_2021}. It's easier to show
this (without loss of generality) for the nearest neighbour Ising
model whose action is given by:
\begin{equation}
S[\phi]=-\beta J\sum_{\mu=1}^{d}\sum_{\bm{r}}\phi(\bm{r}-\hat{\mu})\phi(\bm{r})\label{eq:ising_action}
\end{equation}
where $\phi(\bm{x})$ takes values $\pm1$. We'll assume \emph{open
}boundary conditions for now and relax it later on. Restating the
Boltzmann distribution for the Ising model as an autoregressive process\footnote{$k-\hat{\mu}$ should be understood as the lattice position $\bm{x}-\hat{\mu}$
where $\bm{x}$ maps to $k$ according to the given ordering}:
\begin{equation}
\prod_{k=1}^{N}p(\phi_{k}|\phi_{<k})=p(\phi)=\exp\left(-\beta J\sum_{\mu=1}^{d}\sum_{k}\phi_{k}\phi_{k-\hat{\mu}}\right)/Z\label{eq:log_autoreg}
\end{equation}
From Bayes theorem, we can relate this conditional probability to
the unconditional joint probabilities of the first $k$ and $k-1$
spins, which can in turn be written as integrated forms of the Boltzmann
distribution:
\[
p(\phi_{k}|\phi_{<k})=\frac{p(\phi_{1},\dots\phi_{k})}{p(\phi_{1},\dots\phi_{k-1})}=\frac{{\displaystyle \sum_{\phi_{k+1}\dots\phi_{N}}}p(\phi)}{{\displaystyle \sum_{\phi_{k}\dots\phi_{N}}}p(\phi)}
\]
Expanding the $p(\phi)$ for the Ising model:
\begin{align*}
p(\phi_{k}|\phi_{<k}) & =\frac{{\displaystyle \sum_{\phi_{N},\dots\phi_{k+1}}}\exp\left(-\beta J{\displaystyle \sum_{l=k}^{N}}\left(\phi_{l}{\displaystyle \sum_{\mu}}\phi_{l-\hat{\mu}}\right)+\delta(\phi_{<k})\right)}{{\displaystyle \sum_{\phi_{N},\dots\phi_{k}}}\exp\left(-\beta J{\displaystyle \sum_{l=k}^{N}}\left(\phi_{l}{\displaystyle \sum_{\mu}}\phi_{l-\hat{\mu}}\right)+\delta(\phi_{<k})\right)}
\end{align*}
Since the values in $\phi_{<k}$ are fixed and not summed over, the
terms $\delta(\phi_{<k})$ containing only them cancel from both the
numerator and denominator, leaving us with:
\begin{align}
p(\phi_{k}|\phi_{<k}) & =\frac{{\displaystyle \sum_{\phi_{N},\dots\phi_{k+1}}}\exp\left(-\beta J{\displaystyle \sum_{l=k}^{N}}\phi_{l}{\displaystyle \sum_{\mu}}\phi_{l-\hat{\mu}}\right)}{{\displaystyle \sum_{\phi_{N},\dots\phi_{k}}}\exp\left(-\beta J{\displaystyle \sum_{l=k}^{N}}\phi_{l}{\displaystyle \sum_{\mu}}\phi_{l-\hat{\mu}}\right)}\label{eq:ising_k_conditional}
\end{align}
Even though $\phi_{<k}$ contains $k-1$ values, the conditional probability
$p(\phi_{k}|\phi_{<k})$ depends only on those positions within $\phi_{<k}$
that are nearest neighbours of the positions in $\phi_{\geq k}$.
We can draw the same conclusion for scalar lattice field theory by
replacing the sums with integrals and including terms like $\phi_{l}^{2}$
and $\phi_{l}^{4}$ in the above expression. The number of elements
in the dependency set is bounded above by $L^{d-1}$ or $N/L$ for
our choice of ordering (see figure \ref{fig:open-single-pos} for
an illustration on a $10\times10$ 2D lattice) which is orders of
magnitude smaller than the original upper bound $N$. In fact, we
can join the 2 strips of blue spins in figure \ref{fig:open-single-pos}
into a single 1D line of length $L^{2-1}=10$, and the conditional
distribution on $\phi_{k}$ simply depends on the values along this
line.

We can draw the same conclusion for scalar lattice field theory by
replacing the sums with integrals and including terms like $\phi_{l}^{2}$
and $\phi_{l}^{4}$ in \ref{eq:ising_k_conditional}. Let's split
the lattice vector $\bm{r}$ into a $d-1$ dimensional \emph{spatial
}vector $\bm{x}=[x_{1},\dots x_{d-1}]$ and \emph{time} $t$ so that
$\bm{r}=[r_{1}=x_{1},\dots r_{d-1}=x_{d-1},r_{d}=t]$. In figure \ref{fig:open-single-pos}
for a $10\times10$ lattice, the dependency set of blue positions
is them simply a line of length $L$ along the spatial dimension $x$.
More generally, the dependency set of $\phi_{k(\bm{r})}$ can be cast
as a $d-1$ dimensional spatial sub-lattice that's defined parametrically
using:
\begin{equation}
B_{\bm{r}}^{o}=\left\{ \bm{r}'|\bm{r}'=\begin{cases}
[\bm{x}',t] & \text{if }k([\bm{x}',t])<k(\bm{r})\\{}
[\bm{x}',t-1] & \text{if }k([\bm{x}',t])>k(\bm{r})
\end{cases}\right\} \label{eq:dep_box}
\end{equation}
We'll call this the \emph{dependency surface} at $\bm{r}$. From \ref{eq:ising_k_conditional},
we can write down the joint conditional probabilities for more than
one variable. For example, we can write down this $m$ variable joint
distribution for the Ising model:
\begin{align}
p\left(\left\{ \phi_{k},\dots\phi_{k+m-1}\right\} |\phi_{<k}\right) & =p(\phi_{k}|\phi_{<k})\dots p(\phi_{k+m-1}|\phi_{<k+m-1})\nonumber \\
 & =\frac{{\displaystyle \sum_{\phi_{N},\dots\phi_{k+m}}}\exp\left(-\beta J{\displaystyle \sum_{l=k}^{N}}\left(\phi_{l}{\displaystyle \sum_{\mu}}\phi_{l-\hat{\mu}}\right)\right)}{{\displaystyle \sum_{\phi_{N},\dots\phi_{k}}}\exp\left(-\beta J{\displaystyle \sum_{l=k}^{N}}\left(\phi_{l}{\displaystyle \sum_{\mu}}\phi_{l-\hat{\mu}}\right)\right)}\label{eq:m-conditional}
\end{align}
For the choice of $\bm{r}=[1,\dots1,t]$ and $m=L^{d-1}$, the set
$\left\{ \phi_{k},\dots\phi_{k+m-1}\right\} $ is the \emph{constant
time sub-lattice} $\{\bm{r}'|t'=t\}$, the dependency surface is the
constant time manifold at the previous time step:
\begin{equation}
B_{t}^{o}=\left\{ \bm{r}'|t'=t-1\right\} \label{eq:time-dependency}
\end{equation}
See figure \ref{fig:open-line-pos} for an example in a $10\times10$
lattice. This essentially reduces the overall autoregressive process
to a first order Markov process in the time dimension.

\paragraph*{Periodic boundary conditions}

\begin{figure}
\includegraphics{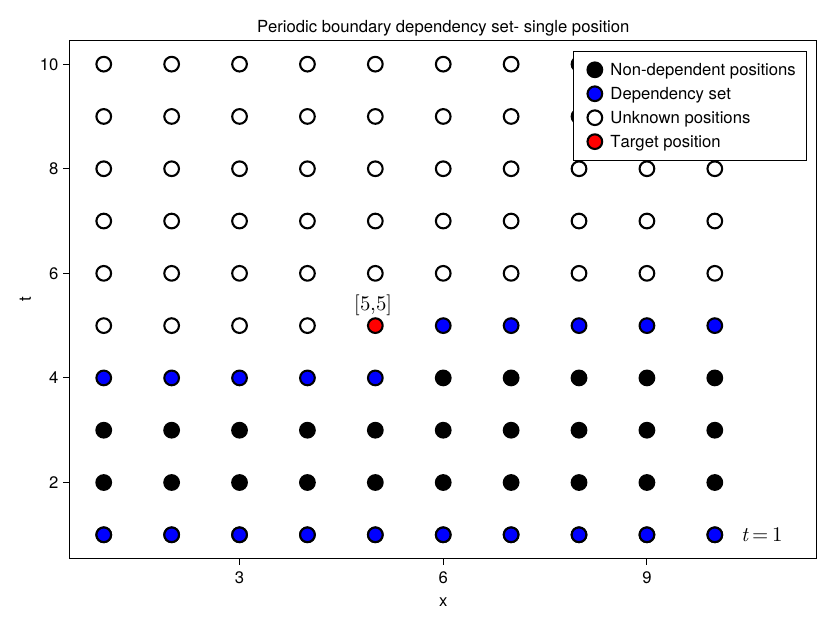}

\includegraphics{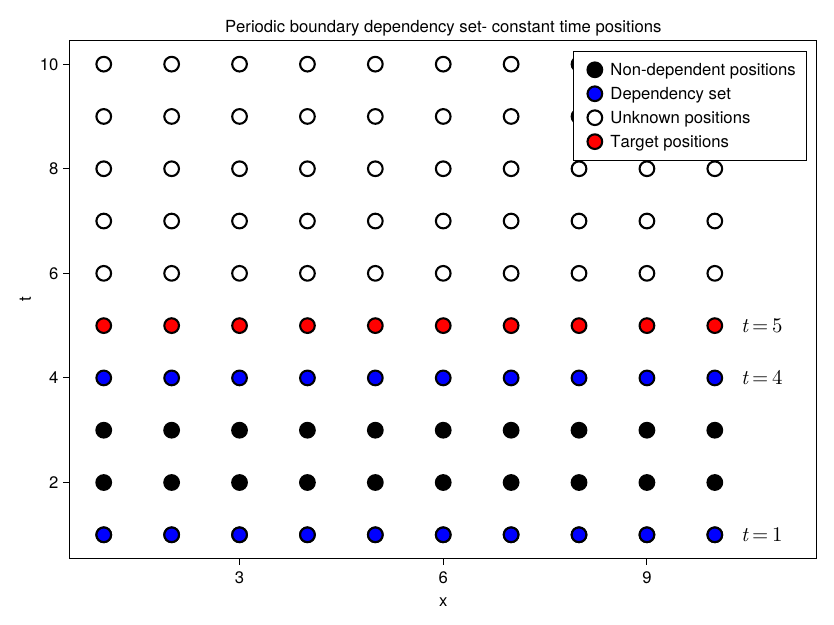}\caption{\label{fig:periodic_figure}In case of a periodic boundary, distributions
of field values at a single red position (top) or positions in a constant
time sub-lattice (bottom) are conditioned on the dependency sets (marked
blue) that are similar to that of an open boundary, but including
the initial time sub-lattice at $t=1$.}
\end{figure}

The general observation that the dependency surface of $\phi_{k(\bm{r})}$
contains only the nearest neighbours of $\phi_{\geq k}$ within $\phi_{<k}$
is true regardless of boundary conditions as long as the action is
local. However, the dependency surface consists of two $n-1$ dimensional
sub-lattices in case of periodic boundary conditions:
\begin{equation}
B_{t}^{p}=B_{t}^{o}\cup\left\{ \bm{r}|r_{d}=1\right\} \label{eq:periodic_dep_set}
\end{equation}
which includes the sub-lattice corresponding to the \emph{initial
time} sub-lattice. Note that periodic boundaries topologically renders
the lattice as $d$ dimensional torus, hence $B_{t}^{o}=\{r_{d}=t-1\}$
would be a $d-1$ torus along the spatial dimensions. This torus effectively
``blocks'' the influence of all equal time $r_{d}<t$ tori except
for the initial one $r_{d}=1$ which is connected to $r_{d}>t$ on
the other side. This is illustrated for $d=2$ in the figure \ref{fig:periodic_figure}.
In other words, only the sub-lattices at $r_{d}=t-1$ and $r_{d}=1$
are \emph{connected} to the sub-lattice $r_{d}=t$ either directly
or via unknown/future positions.

We'll assume periodic boundary conditions in our models since we can
take advantage of the translational invariance of the action\footnote{Circular permutation invariance is the accurate description for finite
lattices with periodic boundary condition, while true translational
invariance strictly occurs only for an infinite lattice.}.

\subsection{Autoregressive Conditional Normalizing Flows and Metropolis sampling\label{subsec:L-ACNF-metropolis}}

\subsubsection{Conditional masked normalizing flow}

Conditional normalizing flow\cite{conditional-norm-flow} models a
bijective map $f_{\theta}$ between samples of a standard prior distribution
$z\sim r(z)$ to samples of a variational distribution conditioned
on $\chi$, $\tilde{\phi}\sim\tilde{q}_{f,\theta}(\tilde{\phi}|\chi)$,
so that $\tilde{\phi}=f_{\theta}(z)$. From the change of variables
formula for probability distributions, we can obtain the log PDF of
the variational distribution:
\begin{align}
\log r(z) & =\log\tilde{q}_{f,\theta}(f_{\theta}(z)|\chi)+\log\left|\det\frac{\partial f_{\theta}(z)}{\partial z}\right|\nonumber \\
\implies\log\tilde{q}_{f,\theta}(f_{\theta}(z)|\chi) & =\log r(z)-\log\left|\det\frac{\partial f_{\theta}(z)}{\partial z}\right|\label{eq:cnf-log-pdf}
\end{align}
The map $f_{\theta}$ is constructed using a sequence of $n$ coupling
layers $g^{(i)}$ whose inputs are acted alternatively by complementary
binary masks $m$ and $1-m$\footnote{All products are element-wise unless stated otherwise}:
\begin{align}
f_{\theta}(z) & =\left(g_{1}\circ\dots\circ g_{n}\right)(z)\nonumber \\
g_{i}(z_{i},\chi) & =m^{(i)}(z_{i})+(1-m^{(i)})\left(\tilde{\phi}e^{s^{(i)}(m^{(i)}(z_{i}),\chi)}+t^{(i)}(m^{(i)}(z_{i}),\chi)\right)\nonumber \\
m^{(i)} & =\begin{cases}
m & i\text{ even}\\
1-m & i\text{ odd}
\end{cases}\label{eq:cnf-sample}
\end{align}
where $z_{1}=z$ and $z_{i}=g_{i-1}(z_{i-1},\chi)$ for $i>1$, so
$z_{n+1}=\tilde{\phi}$. The functions $s^{(i)}$ and $t^{(i)}$ are
in turn modeled using feed-forward neural networks. The log determinant
in \ref{eq:cnf-log-pdf} then becomes:
\[
\log\left|\det\frac{\partial f_{\theta}(z)}{\partial z}\right|=\sum_{i=1}^{n}\log\left|\det\frac{\partial g_{i}(z_{i})}{\partial z_{i}}\right|=\sum_{i=1}^{n}\text{sum}(s^{(i)}(m^{(i)}(z_{i}),\chi))
\]
where the sum function adds up an input array of values. The couplings
can be inverted and composed together for the inverse flow:
\begin{align}
g_{i}^{-1}(z_{i+1},\chi) & =m^{(i)}(z_{i+1})+(1-m^{(i)})\left(z_{i+1}+t^{(i)}(m^{(i)}(z_{i+1}),\chi)\right)e^{-s^{(i)}(m^{(i)}(z_{i+1}),\chi)}\nonumber \\
f_{\theta}^{-1}(\tilde{\phi}) & =\left(g_{n}^{-1}\circ\dots\circ g_{1}^{-1}\right)(\tilde{\phi})\label{eq:inv-cnf-sample}
\end{align}
which is useful to evaluate the log PDF of a known sample $\tilde{\phi}$
of the variational distribution:
\begin{equation}
\log\tilde{q}_{f,\theta}(\tilde{\phi}|\chi)=\log r(f_{\theta}^{-1}(\tilde{\phi}))-\sum_{i=1}^{n}\text{sum}(s^{(i)}(m^{(i)}(z_{i+1}),\chi))\label{eq:inv-cnf-logpdf}
\end{equation}

\subsubsection{Autoregressive process}

We split the Boltzmann distribution as a product of conditional distributions
of the field at single positions in \ref{eq:autoregressive}. We then
generalized this and evaluated the dependency sets for joint conditional
distribution of the field values in a constant time sub-lattice in
\ref{eq:time-dependency}. Let's rewrite this joint conditional distribution
for periodic boundary conditions, but this time conditioned on the
coupling $\lambda$ in \ref{eq:latticeFT-action} as well:
\begin{align*}
p\left(\{\phi(\bm{r})|t\}|\phi_{<[1,\dots t]},\lambda\right) & =p\left(\{\phi(\bm{r})|t\}|\{\phi(\bm{r}')|t'=t-1,1\}\cup\lambda\right)\\
\implies\log p(\phi|\lambda) & =\sum_{t=1}^{L}\log p\left(\{\phi(\bm{r})|x_{d}=t\}|\{\phi(\bm{r})|x_{d}=t-1,1\}\cup\lambda\right)
\end{align*}
where the second line is the full Boltzmann distribution in logarithmic
form. We use the output variational distribution from a conditional
normalizing flow $\tilde{q}_{\theta}(\tilde{\phi}|\chi,t)$ to model
the individual conditional distributions in the RHS, where $\chi=\{\phi(\bm{r})|x_{d}=t-1,1\}\cup\lambda$.
We sample one constant time manifold of $\tilde{\phi}(t)=\{\phi(\bm{r})|x_{d}=t\}$
at a time and supply them as inputs to subsequent calls of the normalizing
flow:
\begin{align}
\tilde{\phi}(1) & \sim\tilde{q}_{\theta}(\tilde{\phi}|\Theta,\Theta,\lambda,1)=\tilde{q}_{\theta}(1)\nonumber \\
 & \vdots\nonumber \\
\tilde{\phi}(k) & \sim\tilde{q}_{\theta}(\tilde{\phi}|\tilde{\phi}(k-1),\tilde{\phi}(1),\lambda,k)=\tilde{q}_{\theta}(k)\nonumber \\
 & \vdots\nonumber \\
\tilde{\phi}(L) & \sim\tilde{q}_{\theta}(\tilde{\phi}|\tilde{\phi}(L-1),\tilde{\phi}(1),\lambda,L)=\tilde{q}_{\theta}(L)\label{eq:autoreg-sampling}
\end{align}
where $\theta$ are neural network weights and $\Theta$ are junk/missing
values\footnote{We supply zero fields along with a boolean flag for the initial sampling.
See code implementation for details.}. Note that we explicitly suppy time as an input since the conditional
distributions aren't necessarily time independent/invariant. Finally,
we can concatenate the samples along the time direction and add the
log PDFs to obtain the sample and corresponding log PDF of the final
model distribution $q_{\theta}(\phi)$ of l-ACNF:
\begin{align*}
\phi & =\text{cat}_{t=1}^{L}\tilde{\phi}(t)\\
\log q_{\theta}(\phi) & =\sum_{t=1}^{L}\log\tilde{q}_{\theta}(t)
\end{align*}
We optimize the weights $\theta$ such that the KL-divergence between
the distributions $q_{\theta}$ and $p$ is minimized:
\begin{align}
\text{KL}(p\bigparallel q_{\theta}) & =\mathbb{E}_{q_{\theta}(\phi)}\left[\log q_{\theta}(\phi)-\log p(\phi)\right]\nonumber \\
 & =\frac{1}{M}\sum_{i=1}^{M}\left[\log q_{\theta}(\phi^{(i)})+S[\phi^{(i)}]\right]\qquad\phi^{(i)}\in\{\phi\sim q_{\theta}\}^{M}\label{eq:KL-loss}
\end{align}
where we sample mini-batches of $\phi$ (of size $M$), evaluate the
mean of the quantity in square brackets and use the gradient of this
quantity wrt $\theta$ to update the weights at every step. We ignore
the log of partition function $\log Z$ in the expansion of $\log p(\phi)$
since it's a constant independent of $\phi$.

\subsubsection{Checkerboard masks and convolutional networks\label{subsec:Scalable-network}}

Particular choices of neural network architecture for $s^{(i)}$and
$t^{(i)}$, and the binary mask $m$ used in \ref{eq:cnf-sample}
have algorithmic benefits in the context of generative modeling of
the scalar lattice field theory. Neural networks containing only $d-1$
dimensional convolutional layers with circular padding (along with
activations and residual connections) and checkerboard mask for $m$
enable a broken translational invariance along the $d-1$ spatial
dimensions of the lattice\cite{multimodal-flow} (as long as it's
a symmetry of the prior distribution $r(z)$ in \ref{eq:cnf-log-pdf}).
No such invariances exist explicitly along the time dimension since
we sequentially sample along this direction, though a well-trained
model is expected to approximately satisfy them. Fully convolutional
networks also enable sampling configurations independent of the lattice
size $L$ along the spatial dimensions. Autoregressive sampling in
\ref{eq:autoreg-sampling} features dependency sets independent of
$L$ along the time dimension, hence we can sample lattices of different
sizes compared to the $L$ that was used to train the model- making
it \emph{scalable}. $L$ is still constrained to be an even integer
in order to avoid ``leaks'' in the normalizing flow due to checkerboard
masks.

For benchmarking our model, we also consider a masked normalizing
flow with $d$ dimensional convolutional layers and circular padding
where the NF generates entire lattice configurations (instead of constant
time sub-lattices) conditioned only on the coupling $\lambda$, similar
to the implementation described in \cite{multimodal-flow}. This admits
translational invariance along all $d$ directions and scalable as
well. l-ACNF has the (theoretical) advantage of better tackling the
\emph{curse of dimensionality} with smaller effective input size ($O(L^{d-1})$)
compared to the latter ($L^{d}$).

\subsubsection{Metropolis-Hastings sampling}

Any bias in the variational distribution after training to minimize
the KL-divergence in \ref{eq:KL-loss} results in bias for various
observables/estimators evaluated after sampling from it. To get unbiased
samples of $p$ from $q_{\theta}$, we run its samples through a Metropolis-Hastings
procedure- a Markov Chain Monte Carlo (MCMC) technique. Starting from
an initial sample $\phi^{(1)}$ and corresponding log PDF $\log q_{\theta}(\phi^{(1)})$,
we update the chain in the $i$th step using current sample $\phi^{(i)}$
and the proposed sample $\phi'$ using:
\begin{align}
A(\phi^{(i-1)},\phi') & =\min\left(1,\exp\left(\log q_{\theta}(\phi^{(i-1)})-\log p(\phi^{(i-1)})-\log q_{\theta}(\phi')+\log p(\phi')\right)\right)\nonumber \\
 & =\min\left(1,\exp\left(\log q_{\theta}(\phi^{(i-1)})-S[\phi^{(i-1)}]-\log q_{\theta}(\phi')+S[\phi']\right)\right)\nonumber \\
\phi^{(i)} & =\begin{cases}
\phi' & A<\pi\\
\phi & \text{else}
\end{cases}\qquad\text{where }\pi\sim\text{Uniform}(0,1)\label{eq:metropolis-hastings}
\end{align}
A necessary condition for the asymptotic convergence of Metropolis-Hastings
samples to $p$ is \emph{ergodicity} which requires:
\[
q_{\theta}(\phi)>0\quad\text{or}\quad\log q_{\theta}(\phi)>-\infty\quad\forall\text{\ensuremath{\phi}}\in\mathbb{R}^{L^{d}}
\]
The normalizing flow in $q_{\theta}$ satisfies this condition if
the prior distribution of the flow satisfies $r(z)>0$ $\forall z\in\mathbb{R}$
which is true in case of a Gaussian prior, for example.

\subsubsection{Symmetrizing and adiabatic retraining}

In the Metropolis-Hastings update step \ref{eq:metropolis-hastings},
it's useful to ensure various symmetry transformations of the proposed
configuration (that render the action $S$ invariant) leave the acceptance
probability $A$ unchanged. This includes translations and reflections
of the lattice, translations along every dimension and the negation
$\phi\rightarrow-\phi$. This amounts to averaging over the PDF $\text{Avg}_{\mathfrak{g}}q_{\theta}(\mathfrak{g}(\phi))$
after applying the transformations $\mathfrak{g}$ and choosing a
randomly transformed field at every step $\mathfrak{g}(\phi)$:
\begin{align}
\log q_{\theta}(\phi) & \rightarrow\text{logsumexp}_{\mathfrak{g}\in G}\left(\log q_{\theta}\left(\mathfrak{g}(\phi)\right)\right)-\log\left|G\right|\nonumber \\
\phi & \rightarrow\mathfrak{g}(\phi)\qquad\mathfrak{g}\sim G\label{eq:symmetrize-1}
\end{align}
where $G$ is the set of all symmetry transformations we consider.
This is one way of tackling poor convergence when the distribution
$p(\phi)$ has well-separated modes as studied in \cite{multimodal-flow}.
In our case, rotations between the time and spatial dimensions also
address the inherent asymmetrical treatment between spatial and time
dimensions as well as lack of invariances along time dimension in
l-ACNF, as mentioned in \ref{subsec:Scalable-network}. The scalability
of the model to sample lattice sizes different from the value it trained
on allows us to perform adiabatic retraining- where we initially train
with smaller $L$ and use this network as an initial state for training
larger $L$, similar to \cite{gauge-equivariant}.

\section{Numerical experiments\label{sec:Numerical-experiments}}

\subsection{Network architecture and training}

The python deep learning library Pytorch\cite{paszke2019pytorch}
was used to build all models used in this study. We model the $\phi^{4}$
theory in a 2 dimensional lattice ($d=2$) with length 16 ($L=16$).
For Model 1, the conditional normalizing flow network contains $n=8$
coupling layers described in \ref{eq:cnf-sample}. As mentioned in
\ref{subsec:Scalable-network}, we use 1D checkerboard binary masks
$m$ and 1D gated convolutional networks\cite{pixelcnn} for the functions
$s^{(i)}$ and $t^{(i)}$, where the conditional variables $\chi$
are processed by a common chain of 4 gated-convolutional networks
and passed as a skip connection to another gated-convolutional network
with input $m^{(i)}z_{i}$. The model is trained using the KL-divergence
loss in \ref{eq:KL-loss} after setting $m^{2}=-4$ and $\lambda\sim$Uniform$(4.8,5.4)$
is sampled at every step- for 9000 steps using Adam optimizer and
learning rate $\eta=10^{-3}$ with Step schedule for $\approx90$
minutes. Model 2 was constructed by adiabatic-retraining Model 1,
with $L=20$ for $\approx40$ minutes using $\eta=10^{-5}$ with the
same optimizer and schedule. Model 3 is constructed using a 2D masked
normalizing flow as described in the second paragraph of \ref{subsec:Scalable-network}
for benchmarking Model 1. It uses 2D checkerboard masks and 2D gated
convolutional networks with circular padding for $n=12$ coupling
layers. This ensures equal number of convolutional operations and
non-linear activations across all models, with slightly higher number
of weights due to 2D convolutional layers. Model 3 was trained with
$L=16$ for $\approx90$ minutes using $\eta=10^{-3}$ to maintain
equivalence with Model 1.

All training and inference was done using an Nvidia RTX 2060 mobile
GPU. Source code is available at \url{https://github.com/dinesh110598/l-ACNF-scalar}.
The data and code for all plots is available in an interactive Julia
notebook here: \url{https://github.com/dinesh110598/l-ACNF-scalar/blob/main/Figures/notebook.jl}.

\subsection{Parameters and observables}

The parameters $m^{2}$ and $\lambda$ in \ref{eq:latticeFT-action}
were chosen close to the $m_{p}L\approx4$ critical line at the symmetrical
phase of the system, similar to the study in \cite{albergo2019flow}.
The difference is we use a single trained model to evaluate all observables
at different lattice lengths $L$ and couplings $\lambda$. Observables
of interest include the 2-point Green's function:
\[
G_{c}(\bm{r})=\frac{1}{N}\sum_{\bm{r}'}\left(\langle\phi(\bm{r}')\phi(\bm{r}'+\bm{r})\rangle-\langle\phi(\bm{r}')\rangle\langle\phi(\bm{r}'+\bm{r})\rangle\right)
\]
its spatial fourier transform
\[
\tilde{G}_{c}(\bm{p},t)=\frac{1}{L^{d-1}}\sum_{\bm{x}}e^{i\bm{p}.\bm{x}}G_{c}(\bm{x},t)
\]
an estimator of pole mass
\begin{equation}
m_{p}(t)=\text{arccosh}\left(\frac{\tilde{G}_{c}(0,t-1)+\tilde{G}_{c}(0,t+1)}{2\tilde{G}_{c}(0,t)}\right)\label{eq:pole_mass}
\end{equation}
the 2-point susceptibility
\[
\chi_{2}=\sum_{\bm{r}}G_{c}(\bm{r})
\]
and the average Ising energy density
\[
E=\frac{1}{d}\sum_{\mu=1}^{d}G_{c}(\hat{\mu})
\]
We also study an autocorrelation function based on the accept/reject
statistics of the Metropolis-Hastings chain of length $T$:
\begin{equation}
\widehat{\rho(\tau)/\rho(0)}_{acc}=\frac{1}{T-\tau}\sum_{j=1}^{T-\tau}\prod_{i=1}^{\tau}\mathbf{1}_{rej}(i+j)\label{eq:acc_autocor}
\end{equation}
where $\bm{1}_{rej}$ is the identity map of a rejected MH step. Another
set of estimators of the autocorrelation is based on the 2-point observables
we defined above
\begin{equation}
\widehat{\rho(\tau)/\rho(0)}_{\mathcal{O}}=\frac{\frac{1}{T-\tau}\sum_{i=1}^{T-\tau}(\mathcal{O}_{i}-\bar{\mathcal{O}})(\mathcal{O}_{i+\tau}-\bar{\mathcal{O}})}{\frac{1}{T}\sum_{i=1}^{T}(\mathcal{O}_{i}-\bar{\mathcal{O}})^{2}}\label{eq:obs_autocor}
\end{equation}
The integrated autocorrelation time wrt acceptance statistics is given
by
\begin{equation}
\tau_{acc}^{int}=\frac{1}{2}+\lim_{\tau_{\text{max}}\rightarrow\infty}\sum_{\tau=1}^{\tau_{\text{max}}}\frac{\rho_{acc}(\tau)}{\rho_{acc}(0)}\approx\frac{1}{2}+\lim_{\tau_{\text{max}}\rightarrow\infty}\sum_{\tau=1}^{\tau_{\text{max}}}\widehat{\rho(\tau)/\rho(0)}_{acc}\label{eq:int_autocor_time}
\end{equation}
and the dynamical critical exponent $z$ for the above quantity is
fit using
\begin{equation}
\tau_{acc}^{int}(L)\approx cL^{z}\label{eq:crit_exp}
\end{equation}

\begin{figure}
\includegraphics{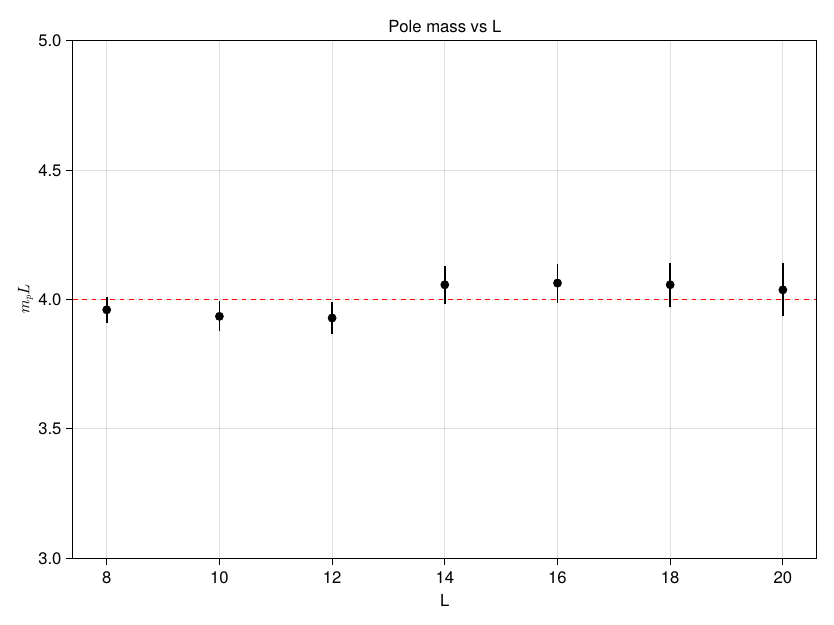}

\caption{Product of pole mass and lattice length $m_{p}L$ plotted against
$L$ evaluated using Model 1. Error bars indicate 68\% confidence
intervals obtained from moving block bootstrap resampling with block
size 100\label{fig:pole_mass}}

\end{figure}

\begin{figure}
\includegraphics{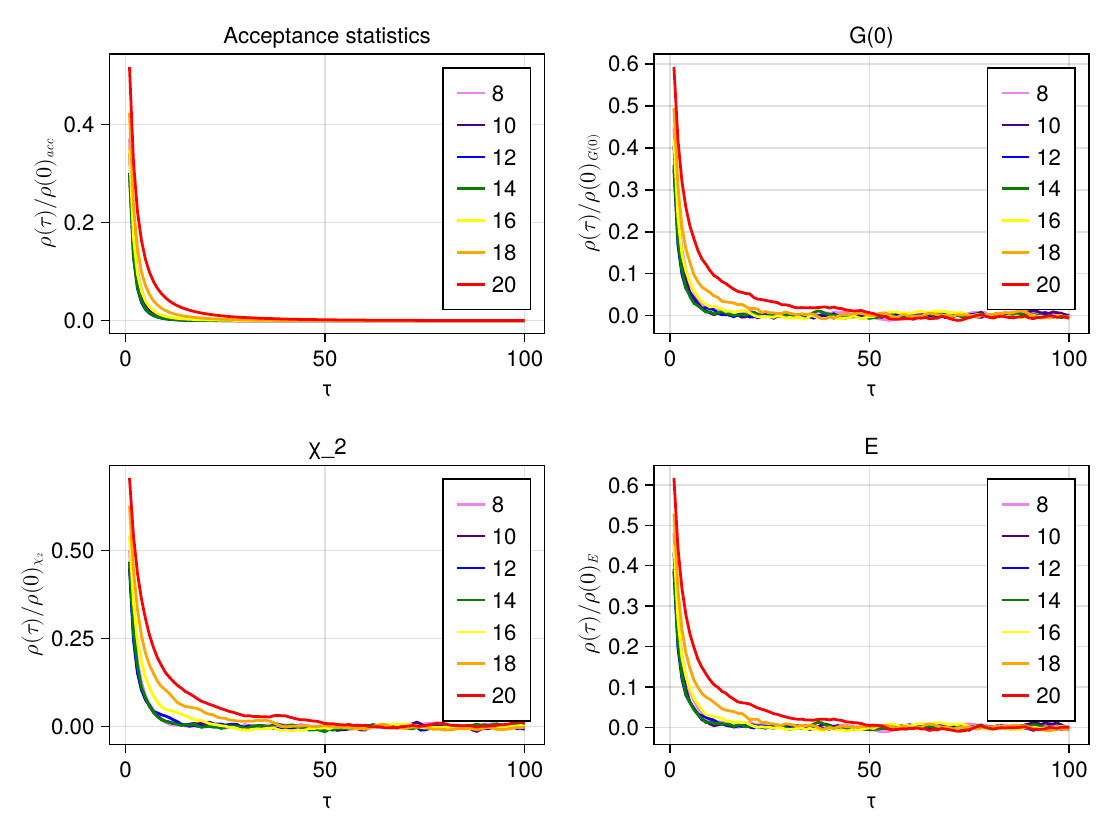}

\caption{\label{fig:autocor_fun} Plots of the autocorrelation functions wrt
acceptance statistics and different observables at various lattice
lengths evaluated using Model 1}

\end{figure}

\begin{figure}
\includegraphics{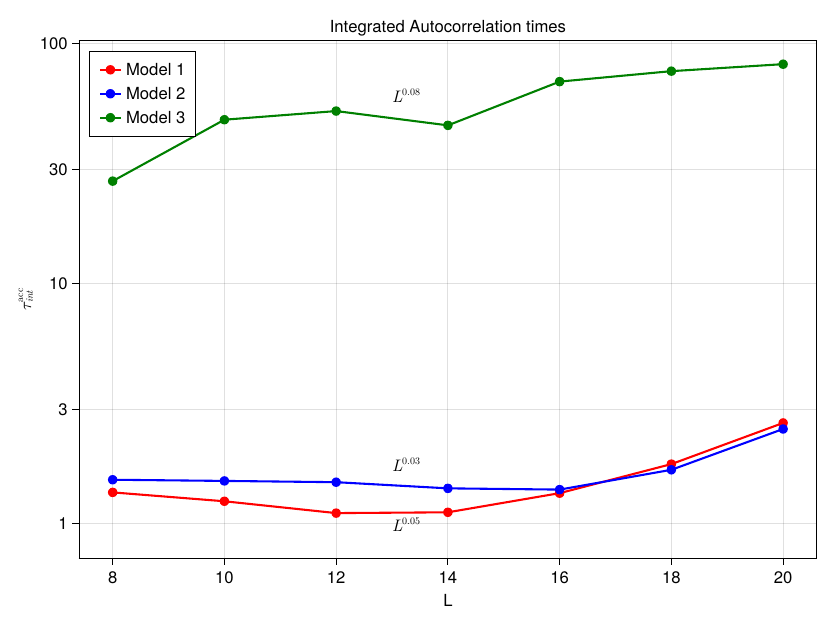}

\caption{\label{fig:int_autocor}Integrated autocorrelation times for acceptance
statistics plotted against $L$ in the logarithm scale using Model
1, 2 and 3}

\end{figure}

\begin{figure}
\includegraphics[scale=0.3]{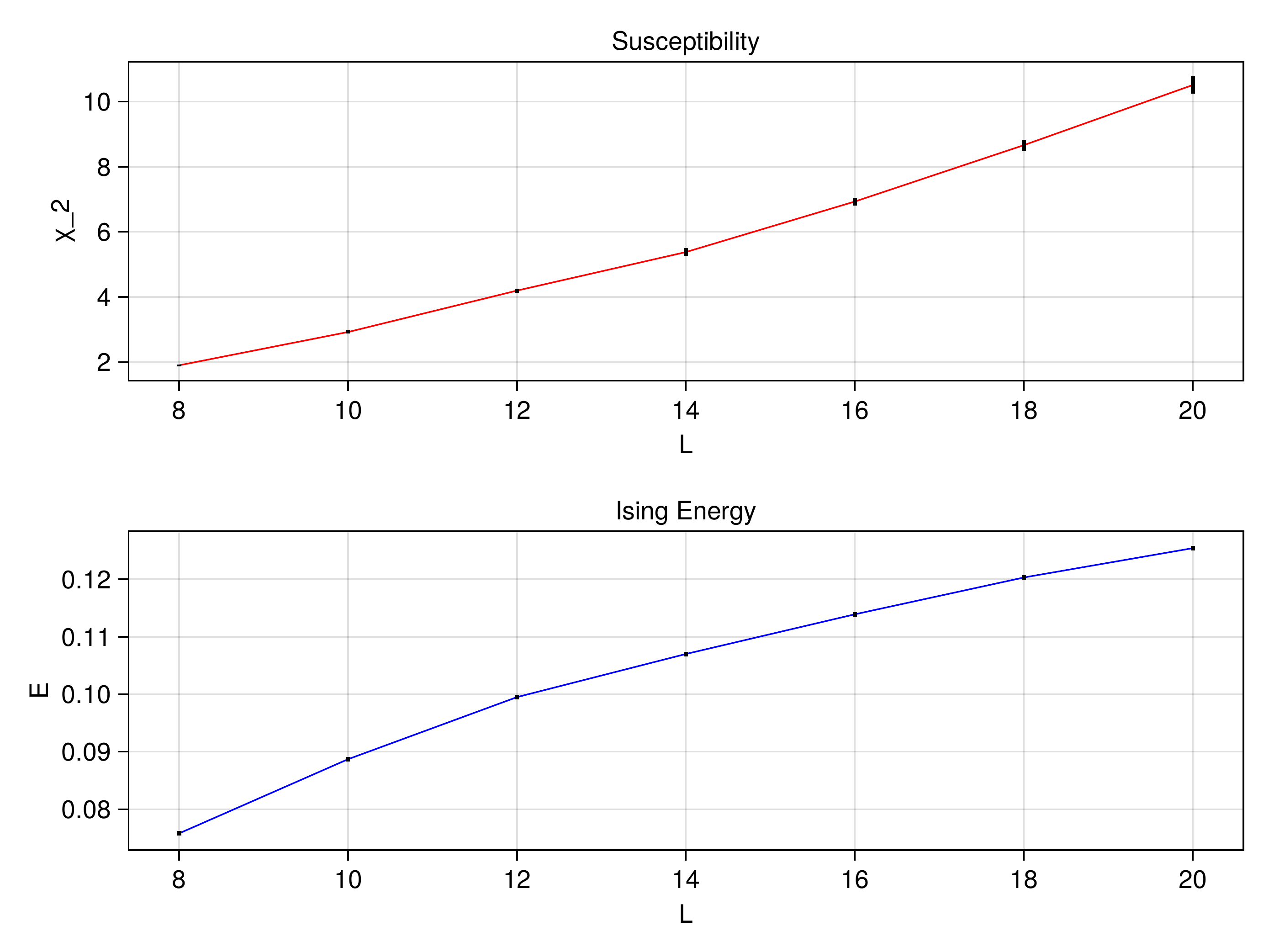}

\caption{\label{fig:obs_L}Plot of observables $\chi_{2}$ and $E$ against
$L$ using Model 1. The error bars indicate 95\% confidence intervals
obtained from moving block bootstrap resampling with block size 100}
\end{figure}

We set $T=10^{5}$ and $L\in[8,10,12,14,16,18,20]$. In every Metropolis-Hastings
step given by \ref{eq:metropolis-hastings}, symmetrization in \ref{eq:symmetrize-1}
was done to transform $\phi^{(i)}$ and $\log q_{\theta}(\phi^{(i)})$
after every ($i$th) step. While this lead to a slower MCMC evaluation,
it significantly improved rates from between $50$-$55\%$ to $65$-$70\%$.
In table \ref{tab:Couplings}, we list the values of $\lambda$ used
in \cite{albergo2019flow} for $L\leq14$ and the remaining fit by
trial and error using Model 1 from \ref{eq:pole_mass}.
\begin{table}
\begin{tabular}{|c|c|c|c|c|c|c|c|}
\hline 
$L$ & 8 & 10 & 12 & 14 & 16 & 18 & 20\tabularnewline
\hline 
\hline 
$\lambda$ & 6.008 & 5.550 & 5.276 & 5.113 & 4.99 & 4.89 & 4.82\tabularnewline
\hline 
\end{tabular}

\caption{Couplings $\lambda$ corresponding to $L\protect\leq14$ were used
in a previous study so that the system lies along the $m_{p}L\approx4$
critical line. Values for $L\protect\geq16$ were determined by trial
and error evaluations of $m_{p}L$ for different $\lambda$ using
our model. See figure \ref{fig:pole_mass} for a plot of $m_{p}L$
vs $L$ at these values of $\lambda$. \label{tab:Couplings}}

\end{table}
Figure \ref{fig:pole_mass} has a plot of $m_{p}L$ vs $L$ at the
specified $\lambda$ and the points are close to the red-dashed critical
line at these points.
 Autocorrelation functions wrt acceptance/rejection statistics (Model
1) and the observables $G_{c}(0),\chi_{2},E$ are plotted based on
\ref{eq:acc_autocor} and \ref{eq:obs_autocor} in figure \ref{fig:autocor_fun}.
 The orange and red curves corresponding to $L=18,20$ seem to decrease
slowly in comparison to others. In the plot integrated autocorrelation
times in figure \ref{fig:int_autocor}, we evaluate $\tau_{int}^{acc}$
from \ref{eq:int_autocor_time} against $L$ using all of models 1,
2 and 3. The most striking feature in this plot is that the green
line representing the autocorrelation times of model 3 has values
that are orders of magnitude larger than that of models 1,2 which
autoregressively sample constant time sub-lattices according to our
proposal. The values for the adiabatically-retrained model 2 in the
blue line is flatter and has a smaller dynamical critical exponent
compared to that model 1 represented by the red line.
 The autocorrelation times generally remain roughly constant between
$L=8$ and $L=14$ and increase between $L=16$ and $L=20$, suggesting
that the quality (in terms of closeness to the actual Boltzmann distribution
$p$) of $q_{\theta}$ drops when sampling lattice sizes larger than
the value $L=16$ it was trained on. This could also be the reason
for growing error bars above $L=16$ of the susceptibility plot (Model
1) in figure \ref{fig:obs_L} as well as for the pole masses in figure
\ref{fig:pole_mass}. Still, the dynamical critical exponents obtained
by fitting with the curve in \ref{eq:crit_exp} are close to zero,
which is usually not the case with traditional methods\footnote{The scaling behavior of flow-based models usually depends on specific
details of the implementation and curve-fitting to an exponential
function can be inaccurate which is evident from the lack of straight-line
behavior in our log-scale plots as well. See the study in \cite{qft-flow-scalability}
for more details.}.

\section{Concluding remarks}

In this work, we proposed a generative model (dubbed l-ACNF) that
autoregressively samples ``equal-time'' sub-lattices of a scalar
lattice field using masked conditional normalizing flows. From the
locality of action $S$ for the $\phi^{4}$ theory in \ref{eq:latticeFT-action},
we determined that these normalizing flows are conditioned on a smaller
sub-lattice of size $O(L^{d-1})$ according to \ref{eq:periodic_dep_set}
instead of the generic upper bound, $L^{d}$. In other words, locality
of the action reduces the autoregressive expansion of the Boltzmann
distribution to a Markov process in time. Models constructed according
to our proposal have reported orders of magnitude smaller autocorrelation
times compared to an equivalently-constructed existing model used
in a previous study in \cite{multimodal-flow}. This is a consequence
of smaller input/output space for the conditional normalizing flows
in our model, essentially addressing the curse of dimensionality in
existing approaches- avoiding exponentially higher number of training
epochs and/or using bulkier neural networks. Constraints in computing
resources limited our study to models that train and converge within
hours on relatively modest hardware. Yet, our models achieve small
autocorrelation times and near-zero dynamic critical exponents similar
to large-scale studies like \cite{albergo2019flow}. If we're working
with the true space-time specification of $d=4$, l-ACNF uses 3D convolutions
for sampling constant time sub-lattices instead of 4D convolutions.
When running on GPU devices, the CUDNN library\cite{cudnn} (used
by any deep learning framework to access GPU-optimized functions/methods)
contains optimized convolution kernels for only upto 3 dimensions
due to their relative practical utility, which offers a slight edge
to l-ANF over alternatives.

Since locality is a generic property of all lattice quantum field
theories, our approach can be extended to more complex theories as
well. For example, the Wilson action for lattice gauge fields
\[
S_{W}=-\sum_{p}\frac{2}{g^{2}}\text{Re}(\text{Tr}(U(p)))
\]
is a sum over plaquettes $p$ which are loops of links $U_{\mu}(\bm{r})$
through nearest neighbours at every position $\bm{r}=[\bm{x},t]$
and along the direction $\mu$. We can extend the analysis in section
\ref{subsec:dep-set-analysis} and determine that the conditional
distribution of links on the equal-time sub-lattice $p\left(\{U_{\mu}(\bm{r})|r_{d}=t\}|\{U_{\mu}(\bm{r})|r_{d}<t\}\right)$
has the dependency set given by (assuming periodic boundary conditions):
\[
B(\bm{r}|r_{d}=t)=\{U_{\mu}(\bm{r})|r_{d}=t-1,\mu<d\}\cup\{U_{\mu}(\bm{r})|r_{d}=1\}
\]
which contains links inside the $r_{d}=t-1$ sub-lattice (except along
the time dimension since they are not part of/connected to plaquettes
containing links along $t=t_{o}$) and the $r_{d}=1$ sub-lattice,
which again reduces the input size of the corresponding flow networks
to $O(L^{d-1})$. Combining our approach with gauge equivariant flows
proposed in \cite{gauge-equivariant1} and \cite{gauge-equivariant}
is an interesting direction to investigate from here, but is unfortunately
beyond the scope of this study.

\printbibliography

\end{document}